\documentstyle[namedreferences,epsf,proceedings]{crckapb}

\begin{opening}

\title{Observations of electromagnetically coupled dust in the Jovian
magnetosphere}

\author{H. Kr\"uger}
\author{E. Gr\"un}
\author{A. Graps}
\author{S. Lammers}
\institute{Max-Planck-Institut f\"ur Kernphysik, \\
Saupfercheckweg 1, 69117 Heidelberg, Germany \\
E-mail: krueger@galileo.mpi-hd.mpg.de}
\end{opening}

\runningtitle{DUST IN THE JOVIAN MAGNETOSPHERE}

\begin{document}

\begin{abstract}
%Measurements of dust particles coupled to the Jovian magnetosphere have been 
%obtained with the dust detector on board the Galileo spacecraft. 
We report on dust measurements obtained during the seventh orbit of the 
Galileo spacecraft about Jupiter. 
%which had a close flyby at Ganymede. 
The most prominent features observed are highly 
time variable dust streams recorded throughout the Jovian system. The 
impact rate varied by more than an order of magnitude with a 5 and 10 hour
periodicity, which shows a correlation with Galileo's position relative to
the Jovian magnetic field. 
%At about 15 $\rm R_J$ distance from Jupiter inbound
%(Jupiter radius, $\rm R_J = 71,492$~km) a reversal in the impact direction by 
%$180^{\circ}$ occurred. 
This behavior can be qualitatively explained by strong 
coupling of nanometer-sized dust to the Jovian magnetic field. In addition 
to the 5 and 10~h periodicities, a longer period which is compatible with 
Io's orbital period is evident in the dust impact rate. This feature indicates 
that Io most likely is the source of the dust streams.
During a close (3,095\,km altitude) flyby at Ganymede on 5 April 1997 an enhanced rate 
of dust impacts has been observed, which suggests that Ganymede is a source of
ejecta particles.  Within a distance of about 25 $\rm R_J$ (Jupiter radius, 
$\rm R_J = 71,492$~km) from Jupiter impacts of 
micrometer-sized particles have been recorded which could be particles on bound 
orbits about Jupiter.
\end{abstract}

\section{Introduction}

In 1992, when the Ulysses spacecraft flew by Jupiter,
the Jovian system was first recognized as a source of intermittent streams of 
submicrometer-sized dust particles (Gr\"un {\it et al.,} 1993). Similar streams  
were later detected within 2 AU distance from Jupiter when Galileo 
was approaching  the planet (Gr\"un {\it et al.,} 1996a). 
It was soon recognized that Jupiter's magnetosphere 
could eject submicron-sized dust particles if they existed in the magnetosphere
(Horanyi {\it et~al.,} 1993a,b; Hamilton and Burns, 1993). 
%Vast amounts of dust had been 
%detected with Voyager's cameras at two places in the Jovian
%system: (1) the Jovian ring at 1.8 $\rm R_J$ (Jupiter radius, 
%$\rm R_J = 71,492\,$ km) and its weak extension out to 3 $\rm R_J$ 
%(Showalter {\it et al.,} 1995), and (2) Io's 
%volcanic plumes (Collins, 1981). Both phenomena have been 
%suggested as the source of the dust streams. 

The arrival direction of the particles detected with both spacecraft 
-- Galileo and Ulysses -- was correlated with the ambient
interplanetary magnetic field which implied electromagnetic interaction 
of the particles forming the dust streams (Gr\"un {\it et al.,} 1993, 1996a).
Only particles with a radius of about 10~nm can couple strongly enough to 
the interplanetary magnetic field to show the observed effects 
(Zook {\it et al.,} 1996; J.C. Liou, priv. comm., 1997). 
The deduced particle impact speeds exceed 200 km/s. 

%On 7 December 1995 Galileo completed its journey to Jupiter and was 
%inserted into a highly elliptical orbit about the planet. 
%During this initial approach to Jupiter and Io
%%, dust continuously impacted 
%%the dust detector on board. After Io closest approach impacts of 
%%small particles ceased (Gr\"un {\it et al.,} 1996b). A few larger 
%%particles were recorded within hours of Jupiter closest approach. 
%data transmission was very limited due to technical constraints, and
%additionally, the sensitivity of the dust sensor 
%was reduced. Both effects resulted in only a small 
%number of recorded impacts which did not allow for a detailed analysis of 
%dust in the inner Jovian magnetosphere. 
%%Shortly after Jupiter flyby, the dust instrument  
%%was deactivated and was only reactivated half a year 
%%later before the first Ganymede flyby.

Being now in a highly elliptical orbit about Jupiter, Galileo performs 
dust measurements in the inner Jovian magnetosphere and 
during close flybys of the Galilean  satellites (Gr\"un {\it et al.}, 1997, 
1998). 
%At least three types of dust particles have been identified in 
%the Jovian system (Gr\"un {\it et al.}, 1997, 1998): (1) small 
%submicron-sized dust particles with high and variable impact rates throughout 
%the Jovian system, (2) a concentration of small dust impacts at the times of 
%closest approaches to the three outer Galilean satellites, and (3) big 
%micrometer-sized dust particles concentrated in the inner Jovian system.  
In this paper we present data from Galileo's seventh orbit about Jupiter 
(G7 orbit) which had a close (3,095\,km altitude) flyby of Ganymede on 
5 April 1997. In the time period of 10 days considered here, Galileo was 
within a distance of 50 $\rm R_J$ from Jupiter. 

\section{The dust detector}

The Galileo dust detector (DDS) is a multicoincidence impact ionization 
detector (Gr\"un {\it et al.,} 1992a; 1995) which measures submicrometer- 
and micrometer-sized dust particles. The instrument has been 
calibrated in the laboratory for impact velocities between 2\, km/s and 
70 km/s. DDS is identical with the dust
detector on board Ulysses.  For each dust impact onto the sensor,
three independent measurements of the impact-created plasma cloud
are used to derive the impact speed and  the mass of the particle. The
coincidence times of the three charge signals are used to classify each
impact. Class 3, our highest class, are real dust impacts, and class 0
are noise events. Class~1 and class~2 events were true
dust impacts in interplanetary space  (Baguhl {\it et al.,}
1993; Kr\"uger {\it et al.,} 1998).
%Within 18$\rm R_J$ distance from Jupiter the event definition status
%which initiates a measurements cycle was changed in
%order to avoid dead-time problems and to reduce the noise sensitivity
%of the channeltron. The sensitivity for class 2 and class 3 dust
%impacts was not affected by this change.
%Although channeltron noise was reduced, energetic particles still
%caused enhanced noise rates 
In the Jovian system, within about 20 $\rm R_J$ from Jupiter, however,
class~1 and class~2 are affected by noise. Noise events could be 
eliminated from the class~2 data (Kr\"uger {\it et al.} 
in prep.), but class~1 events show signatures of
being all noise events in the Jovian environment. 
In this paper, we discuss class 3 and denoised class~2 dust data.
Apart from a missing third charge signal, there is no physical difference 
between dust impacts which are categorized into class~2 and class~3,
respectively. Class~3 impacts have three charge signals, whereas only two 
are required for a class~2 event.

\begin{figure}
\epsfxsize=8.5cm
\epsfbox{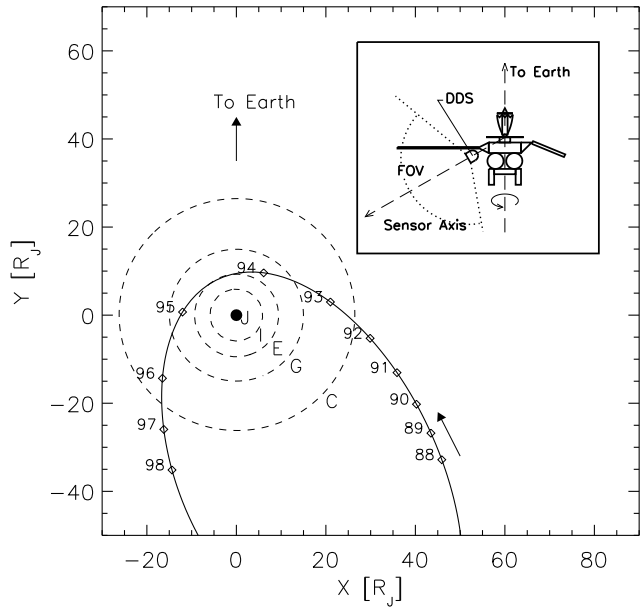}
        \caption{\label{orbit} 
Galileo's orbit trajectory during the Ganymede 7 flyby projected 
onto Jupiter's (J) equatorial plane (Jupiter radius, $\rm R_J = 
71,492\,km$). The orbits of the Galilean satellites are shown: Io (I), 
Europa(E), Ganymede (G), and Callisto (C). Dates are marked by 
diamonds (numbers give day of year) on Galileo's path through the 
Jovian system. Earth direction is to the top. The inset shows a sketch 
of the spacecraft and the field of view (FOV) of the dust detector (DDS).
}
\end{figure}

Galileo is a dual-spinning spacecraft, with an antenna that points
antiparallel to the positive spacecraft spin axis (PSA, cf. inset in 
Fig.~\ref{orbit}). During most of the  
orbital tour around Jupiter, the antenna points towards Earth. 
DDS is mounted on the spinning section of the spacecraft
and the DDS sensor axis is offset by an angle of $60^{\circ}$ from the spin 
axis (an angle of $\rm 55^{\circ}$ has been erroneously
stated before). The rotation angle measures the viewing
direction of the dust sensor at the time of a particle impact. During one spin
revolution of the spacecraft, the rotation angle scans through $360^{\circ}$.
DDS has a $140^{\circ}$ wide field of view. A reduction in the 
field of view has recently been recognized 
for a subset of the smallest dust impacts (Kr\"uger {\em et al.} in prep.) 
which does not apply to the present analysis. Dust particles that arrive
from within $10^{\circ}$ of the PSA can be sensed at all rotation angles,
while those that arrive at angles from $10^{\circ}$ to $120^{\circ}$ from 
the PSA can only be sensed over a limited range of rotation angles.

\section{Dust Measurements from Galileo's G7 orbit}  \label{measurements}

\begin{figure}
\epsfxsize=12.5cm
\epsfbox{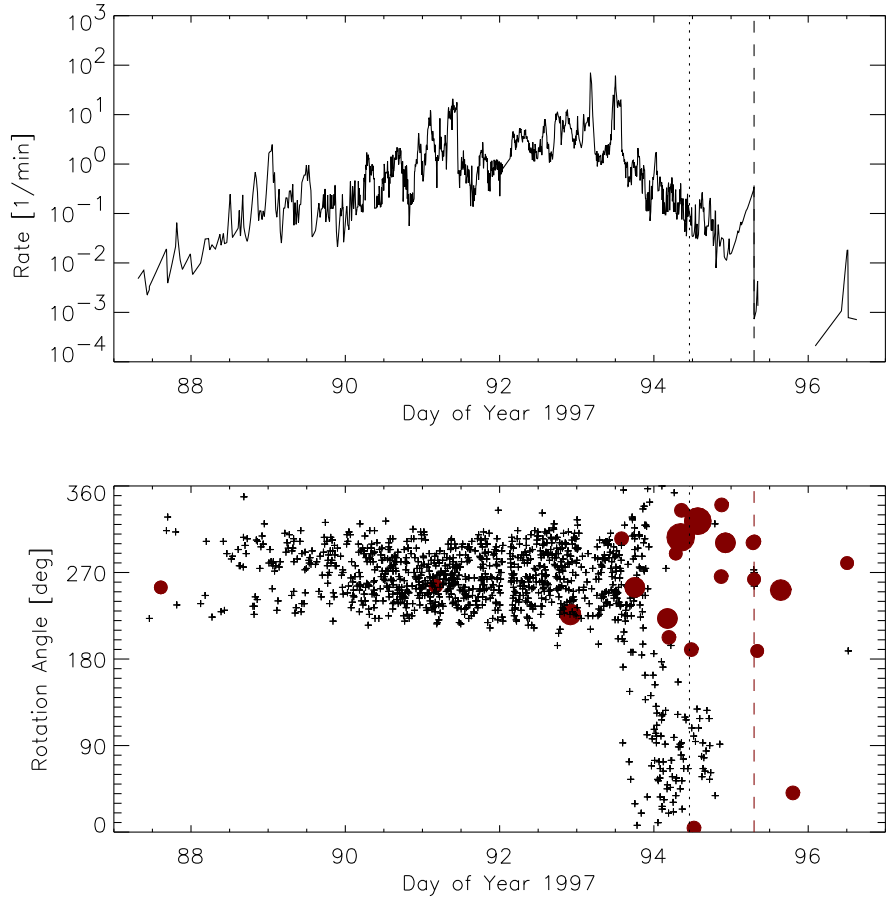}
        \caption{\label{plastro1}
Dust data from Galileo's G7 orbit (28 March to 6 April 1997).
Upper panel: Impact rate of small dust particles (impact charge $\rm Q_{\,I} 
< 10^{-13}\,C$, class~2 and class~3). Lower panel: Sensor direction at time of 
dust particle impact (rotation angle) for those impacts for which the complete 
information has been transmitted to Earth (class~2 and class~3). Small 
particles (impact charge $\rm Q_{\,I} <
10^{-13}\,C$) are shown as crosses, bigger particles are shown as filled circles.
The size of the circles indicates the impact charge ($ \rm 10^{-13}\,C \leq 
 Q_{\,I} \leq 10^{-10}\,C$). At $0^{\circ}$ the sensor points close to the 
ecliptic North direction, at $90^{\circ}$ and $270^{\circ}$ the sensor points 
close to Jupiter's equatorial plane. The closest approach to Ganymede is
indicated by a dashed line and perijove passage by a dotted line. 
}
\end{figure}

During Galileo's approach to Jupiter DDS has detected a highly 
variable impact rate of dust particles. The upper panel of 
Fig.~\ref{plastro1} shows the impact rate of our smallest 
dust impacts in classes~2 and 3 (ion collector charge 
$\rm Q_I < 10^{- 13}$~C) for a period of 10 days. 
In the following we will call these particles dust stream particles.
The closest approach to Ganymede was on 5 April 1997 (day 95). 
On day 88 -- when Galileo was at about 
50$\rm R_J$ distance from Jupiter -- a dust impact rate of about 0.01 
imp/min was detected. Closer to Jupiter the impact rate 
exceeded 10 imp/min a few times (days 91 and 93). Around perigee passage 
the impact rate dropped to about 0.1 imp/min. Between day 
88 and day 94 the impact rate fluctuated by more than an order of 
magnitude with periods of about 5 and 10~hours. 
At Ganymede closest approach a sharp peak occurred which lasted 
only several minutes (Fig.~\ref{plastro1}). 
%Fluctuations with 5 and 10 hour periods 
%and sharp peaks at closest approach to the Galilean satellites were also 
%reported from earlier orbits of Galileo about Jupiter 
%(Gr\"un {\it et al.}, 1998).

The impact direction (rotation angle) of the dust particles as derived from 
the sensor orientation at the time of particle impact is shown in the bottom 
panel of Fig.~\ref{plastro1}. The impact direction of a single particle is
only known to lie somewhere within the $140^{\circ}$ wide sensitive solid
angle cone of DDS. The average of all the rotation angle arrival
directions of dust particles belonging to a stream is known to much
higher accuracy than is that angle for a single particle.

When Galileo was approaching Jupiter the dust impact direction 
(rotation angle) was concentrated between $210^{\circ}$ and $330^{\circ}$. 
Half a day before perijove passage the impact direction of small
particles shifted by 
$180^{\circ}$ and dust particles approached from the opposite direction. 
This change in impact direction is coincident with the drop in the
impact rate  on day 94.0 (cf. Fig.~\ref{plastro1}, upper panel). 
On day 94.8 impacts of small stream particles ceased. 
(One small event on day 96.5 is most likely caused by incomplete 
denoising of class~2).
The vast majority of the particles were small submicrometer-sized dust 
particles which just exceeded the detection threshold 
(impact charges $\rm Q_{\,I} \geq 10^{-14}\,C$). 
Only about 20 bigger particles ($\rm 10^{-13}\,C \leq Q_{\,I} \leq 
10^{-11}\,C$) were detected within two days around perijove passage, i.e.
within 25~$\rm R_J$ from Jupiter.

Particles detected within about an hour of Ganymede closest approach came 
from the  direction between $250^{\circ}$ and $310^{\circ}$. Note that 
the dust streams had already vanished half a day earlier.
 
\section{Discussion} \label{discussion}

Three types of dust particles have been detected within the Jovian 
system (Gr\"un {\it et al.}, 1997, 1998): 
(1) small submicron-sized dust particles with high and variable impact rates 
throughout the Jovian system, (2) a concentration of small dust impacts at the 
times of closest approaches to the Galilean satellites, and (3) big 
micrometer-sized dust particles concentrated in the inner Jovian system.  

The dust streams observed in interplanetary space out to 2~AU from 
Jupiter have been explained as streams of tiny particles electromagnetically
ejected from the Jovian system (Hor\'anyi {\it et al.}, 1993a,b; Hamilton
and Burns, 1993). 
The tiny dust particles we see within the Jovian system with highly
time variable impact rates with periods of about 5 and 10~h (category 1 
above) are the continuation of these dust streams 
detected in interplanetary space. 
Both types of dust particles show the same characteristics in 
terms of impact magnitude and impact direction, and there is 
a smooth transition of the impact rate from interplanetary space to 
within the Jovian magnetosphere (Gr\"un {\it et al.}, 1996b). 
The radii of the particles are around 10~nm and their velocities 
exceed 200~km/s (Zook {\it et al.}, 1996) which is a factor of three 
beyond the upper limit of the instrument's calibrated velocity range.

When Galileo is approaching Jupiter, a 
radial outflow of dust is initially detected from rotation angles 
of $270^{\circ}$. When Galileo moves closer to Jupiter, the impact direction 
moves closer to the anti-Earth direction (cf. Fig.~\ref{orbit}), and dust 
arrives parallel to Galileo's spin axis.
Shortly thereafter, the impact direction changes to rotation 
angles of $90^{\circ}$. After perijove passage, Jupiter, Earth and the dust
source are in the same hemisphere when viewed from Galileo and dust 
particles approach from a direction close to the negative spacecraft spin 
axis (Earth direction). From this direction, dust grains
usually cannot enter the DDS sensor, and remain mostly 
undetected. This explains why DDS has sensed only a few dust particles 
on the outbound part of its orbit. 

\begin{figure}
\epsfxsize=12.5cm
\epsfbox{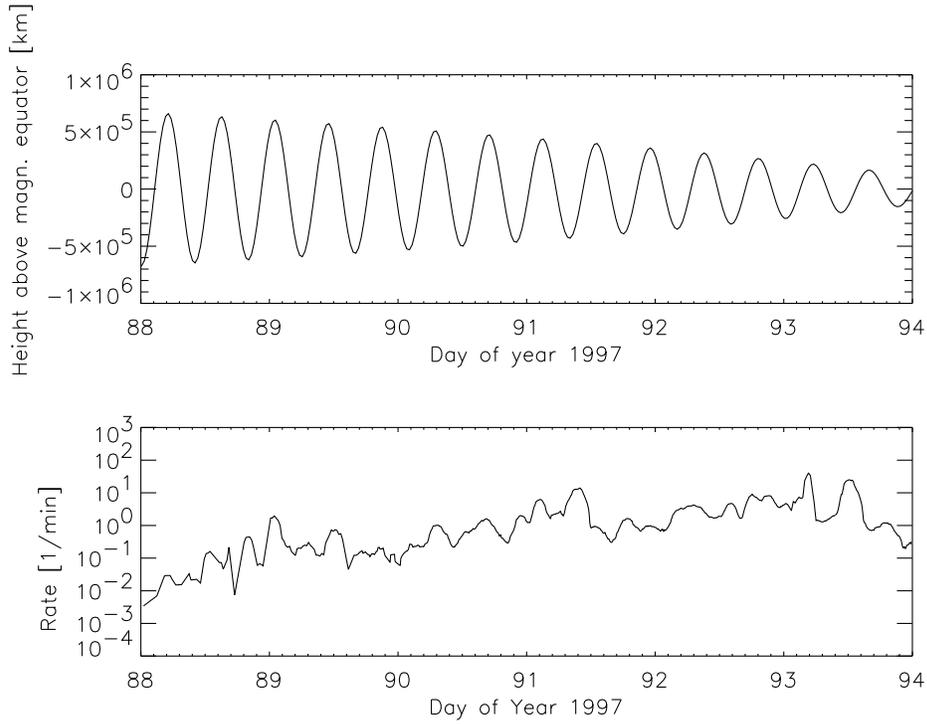}
        \caption{\label{plastro2}
Phase relation between Galileo's position in Jupiter's magnetic field 
(here Galileo's distance from the magnetic equatorial plane is shown) and the
observed impact rate.  A dipole tilted by $9.6^{\circ}$ w.r.t. 
Jupiter's rotation axis pointing towards $\rm \lambda_{III}= 202^{\circ}$ 
has been adopted for the magnetic field. The observed impact rate is 
smoothed with a 2 hour average. Note the 5 and 10~h periodicity in the 
impact rate.
} 
\end{figure}

Fluctuations in the dust impact rate with 5 and 10~h periods 
(Fig.~\ref{plastro1}) have also been reported from other orbits of 
Galileo about Jupiter (Gr\"un {\it et al.}, 1998). In Fig.~\ref{plastro2} we 
study the phase relation between the impact rate and Jupiter's magnetic field at 
the position of Galileo in a subset of the G7 data. The upper
panel shows Galileo's distance from the magnetic equatorial plane 
($\rm Z$~position in Jupiter's magnetic field).
% a tilted dipole field has been adopted). 
The lower panel shows 2 hour averages of the observed impact rate.
The 5 and 10 hour periods are evident in the impact rate which indicates 
that the impact rate is modulated by the magnetic field. Other impact parameters
like the rotation angle, the charge rise times and the charge amplitudes are also 
correlated with the magnetic field position.  This indicates that the 
impact direction, impact velocity, and sizes of the impacting dust 
particles are also modulated by the magnetic field. Fourier analysis of 
the observed impact rates shows peaks in the frequency spectrum
at about 10 hours (Jupiter's rotation period) and at about half 
that period.  No lower frequencies have been found in 
previous data sets  (Gr\"un {\it et al.}, 1998). 

The observed features can be qualitatively explained by a simple model of dust
which is released in the inner Jovian system and electromagnetically coupled 
to the magnetosphere (Gr\"un {\it et al.}, 1998, 
Horanyi {\it et al.}, 1993a,b, Hamilton and Burns, 1993).  Here we 
recall only the most important aspects. 
Particles ejected from a source in the inner Jovian system (Io or the 
gossamer ring) enter the Io plasma torus and are later released from the 
torus with a typical charge of about +3V (Horanyi {\it et al.}, 1997).
Once a particle is positively charged it will be accelerated outward by 
the co-rotational electric field. Because Jupiter's magnetic field is tilted 
by $9.6^{\circ}$ w.r.t. to the planet's rotation axis the particles move in 
a warped dust sheet. The particle trajectories are typically bent out of the 
equatorial plane by 10 to $20^{\circ}$. An observer in 
Jupiter's equatorial plane detects dust particles when this dust sheet
passes over its position. This occurs twice per Jupiter 
rotation and a periodic flux results with a 5~h period which 
explains the observed 5 and 10~h periodicities.
The shift in impact direction by $180^{\circ}$ on day 94.0 is best 
explained by particles with a radius of about $\rm 10\,nm$ which move on 
highly bent trajectories.

The Jovian ring (Showalter {\it et al.}, 1995) and Io (Collins, 1981) 
have been suggested as possible sources of the dust. If Io were the source,
its footprint  should be obvious in the data. 
The impact rate in the G7 data set (Fig.~\ref{plastro1})
shows indeed three large peaks, each about two days apart (days
89, 91 and 93). The times between these peaks are close to Io's orbital 
period (1.77 days). 
For a proper analysis of the phase relation between the dust impact rate and 
Io's orbital period one has to consider the relative motion between Galileo 
and Io. Because Galileo moves on an elliptical orbit, its 
velocity depends on its orbital position.
Therefore, Io's  orbital period as seen from Galileo 
depends on Galileo's distance from Jupiter: far away from Jupiter an 
observer on Galileo sees Io orbiting about Jupiter with nearly exactly 
its orbital period. Close to Jupiter, however, Galileo and Io 
have comparable orbital velocities which leads to 
an expansion of Io's orbital period as seen from Galileo.

\begin{figure}
\epsfxsize=12.5cm
\epsfbox{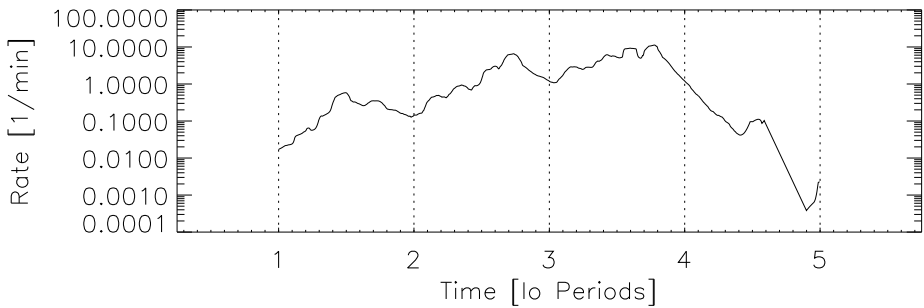}
        \caption{\label{ioperiod}
12~hour average of the dust impact rate as a function of time in units of 
Io's orbital period as seen from Galileo. At each dotted line the distance between 
Io and Galileo has a minimum. 
} 
\end{figure}

Figure~\ref{ioperiod} shows 12 hour averages of the dust impact rate as a 
function of time in units of Io's orbital period as seen by an observer 
moving with Galileo. Periods shorter than 12 hours are suppressed in this 
diagram. At each dotted line the distance between Galileo and Io has a minimum.
Minima in the impact rate occur at minima in the distance from 
Io, and maxima in the impact rate roughly coincide with maxima in the distance.
This indicates that the impact rate of the dust stream particles is 
correlated with distance from Io.
Fourier analysis of a large data set from Galileo's primary mission (1996
and 1997, data from eleven Galileo orbits about Jupiter) also shows a peak 
at Io's period, which was not visible in earlier smaller data sets.

The present results are compatible with Io being the source of the 
dust stream particles. 
In the simplest picture with Io being a point source for dust particles, one
would expect a variation of the impact rate with Io's period. 
The relatively weak modulation of the impact rate 
by Io indicates that Io's period is smeared-out by the Io plasma 
torus which then acts as a more continuous source for dust particles. In this
model one would expect maxima in the impact rate to coincide with 
minima in the distance from the source rather than maxima in the 
distance. This is explained 
by highly bent particle trajectories: particles 
which hit the detector were actually released from the source 
when the source was on the far side as seen from Galileo (cf. Gr\"un {\it 
et al.}, 1998, the trajectory of the 9 nm particle in their Fig. 11).
Analysis of the gravitational and electromagnetic forces indicates that 
particles smaller than about $\rm 10\, nm$ may be able to  overcome Io's 
gravity and escape from the satellite (Ip, 1996).

As noted earlier (Gr\"un {\it et al.}, 1997, 1998), there is a second category 
of particles, which occurred within a few minutes of the closest 
approach (CA) to Ganymede. It has the signature of dust released from Ganymede: 
the impact rate is strongly peaked at the time of CA (day 95),
and the impact direction is compatible with a 
satellite source. Only particles arriving from about $270^{\circ}$ can be 
explained by a Ganymede 
source, whereas $90^{\circ}$ is not compatible with such a source. 
The calibrated velocities of these particles are close to
the relative velocity between Galileo and Ganymede. This
indicates that our empirial calibration of the dust instrument 
(Gr\"un {\it et al.}, 1995) can be applied to these particles which is not 
the case for the stream particles. By adopting the empirical calibration
the size of the particles is about $\rm 0.5\mu m$.
We interpret these as secondary ejecta particles which are generated 
by impacts of other particles onto the surface of Ganymede. Such a 
process has been suggested as being responsible for maintaining both 
the Jovian ring (Morfill {\it et al.}, 1980; Burns, 1980, 
Hor\'anyi and Cravens, 1996) and Saturn's E-rings 
(Hamilton and Burns, 1994). These ejecta particles, however, are 
not the main source of the dust stream particles described above.
A detailed analysis of these secondary particles is forthcoming (Kr\"uger
{\it et al.}, in prep.). 
Secondary particles are also noticeable during encounters with Europa and 
Callisto. 

Impacts of large (micrometer-sized) particles have been recorded mainly in
the inner Jovian system when Galileo was within a distance of about 
$\rm 25\,R_J$ from Jupiter. 
Large-particle impacts behave quite differently from dust 
stream particles (cf. Fig.~\ref{plastro1}): their impact rate peaks 
near perijove passage where the shift of the rotation angle occurs. Beside 
a population of particles orbiting Jupiter on prograde orbits, which
could be ejecta  from the inner Jovian satellites, there has to be 
a population on retrograde orbits (Thiessenhusen {\it et al.}, 1998).
%Large-particle 
%impacts prior to perijove  passage with rotation angles of about $270^{\circ}$ are 
%compatible with particles orbiting on prograde or retrograde orbits, 
%whereas impacts after perijove passage with rotation angles of $90^{\circ}$ can 
%best be explained by highly inclined or even retrograde orbits. 
These particles could originate from Jupiter's satellites 
or could be captured interplanetary and interstellar particles
(Colwell {\it et al.,} 1998).
Such a population of particles has also been deduced from data by 
earlier spacecraft 
carrying dust detectors through the Jovian system: Pioneers 10 and 11 
(Humes {\it et al.} 1974, Humes 1980) and Ulysses (Gr\"un {\it et al.,} 1992b).

\section{Summary}

During Galileo's orbits around Jupiter, the dust detector on board
recorded a highly variable impact rate. In this paper we have 
considered data from Galileo's seventh orbit (G7). 

1. The main signature of the data, namely the 5 and 10 hour periodicities 
of the impact rate are most 
naturally explained by the coupling of charged dust particles to 
Jupiter's magnetic field. 

2. In addition to the 5 and 10 h periods, Io's orbital period (42.4 h)
is recognized in the impact rate data. This result indicates that 
Io is most likely the source of the dust stream particles. Io does not
appear as a strong point source, however, because particles initially
ejected from Io can be dispersed all around the torus before being 
released from the torus and driven out of the Jovian system. This 
leads to a picture of an extended source 
(the torus) with additional point source features (Io itself). A 
more detailed analysis of the data is needed to finally answer this 
question (Graps {\it et al.} in prep.).

%3. The shift in impact direction by $180^{\circ}$ at 20~$\rm R_J$ 
%is a geometric effect. The data are best explained by tiny 
%particles (about $\rm 10 nm$ radius) which move on strongly curved trajectories. 

3. An enhanced impact rate within minutes of closest approach to Ganymede
indicates a population of ejecta particles from this 
satellite. The impact directions and velocities are compatible
with such an origin of the particles.

4. Within a distance of about 25~$\rm R_J$ from Jupiter, impacts of large 
particles have been recorded. Some impacts are 
compatible with particles on prograde orbits, but there has to be a 
significant fraction of particles on retrograde orbits as well.

\vspace{0.5cm}

{\bf Acknowledgements}
We thank the Galileo project at JPL for effective and successful 
mission operations. This work has been supported by Deutsches
Zentrum f\"ur Luft- und Raumfahrt (DLR).

%\begin{quote}
%{\small
%\begin{verbatim}

%\end{verbatim}
%}
%\end{quote}

\end{document}